\RequirePackage{fixltx2e}
\documentclass[12pt]{iopart}

\usepackage[utf8]{inputenc}
\usepackage[brazilian,british]{babel}
\usepackage[T1]{fontenc}
\usepackage{amssymb,amsthm}
\usepackage[sc,osf]{mathpazo}
\usepackage[pdftex]{graphicx}
\usepackage[update,prepend]{epstopdf}
\usepackage{csquotes}
\usepackage{verbatim}
\usepackage{xcolor}
\usepackage{bm}
\usepackage[babel=true]{microtype}
\usepackage{pifont}
\usepackage{multirow}
\usepackage{booktabs}  
\usepackage{iopams}
\usepackage[backend=bibtex,bibencoding=utf8,maxbibnames=99]{biblatex}
\newcommand{\mean}[1]{\left\langle#1\right\rangle}

\newcommand{\norm}[1]{\left\Vert#1\right\Vert}

\newcommand{\trace}[1]{\mbox{tr}\left( #1 \right)}

\newcommand{\ket}[1]{\vert#1\rangle}
\newcommand{\bra}[1]{\langle#1\vert}

\newcommand{\ketbra}[2]{\vert #1 \rangle \langle #2 \vert}
\newcommand{\avg}[1]{\langle#1\rangle}

\newcommand{\eg}{\emph{e.g.}}
\newcommand{\ie}{\emph{i.e.}}
\newcommand{\cf}{\emph{cf. }}

\newcommand{\refe}[1]{(\ref{#1})}

\begin{document}

\title[]{Quantum process tomography with informational incomplete data of two $J$-coupled heterogeneous spins relaxation
in a time window much greater than $T_1$}

\author{Thiago O. Maciel, Reinaldo O. Vianna}
\ead{thiago@fisica.ufmg.br}
\address{Departamento de F\'{\i}sica - ICEx - Universidade Federal de Minas Gerais,
Av. Ant\^onio Carlos 6627 - Belo Horizonte - MG - Brazil - 31270-901.}
\author{Roberto S. Sarthour, Ivan S. Oliveira}
\address{Centro Brasileiro de Pesquisas F\'{\i}sicas, \\
Rua Dr. Xavier Sigaud 150 - Rio de Janeiro - RJ - Brazil - 22290-180.}

\vspace{10pt}
\begin{indented}
\item[]April 2014
\end{indented}

\begin{abstract}
We reconstruct the time dependent quantum map corresponding to the relaxation process of a two-spin system in liquid-state NMR at room temperature. 
By means of quantum tomography techniques that handle informational incomplete data, we show how to properly post-process and normalize the measurements data for the
simulation of quantum information processing, overcoming the unknown number of molecules 
prepared in  a non-equilibrium magnetisation state ($N_j$) by an initial  sequence of radiofrequency (RF) pulses. From the 
reconstructed quantum map, we infer both longitudinal ($T_1$) and transversal ($T_2$) relaxation times, and introduce the $J$-coupling
relaxation times ($T^J_1,T^J_2$), which are relevant for quantum information processing simulations. We show that the map associated 
to the relaxation process cannot be assumed approximated unital and trace-preserving for times greater than $T_2^J$. 
\end{abstract}

%
%
\submitto{\NJP}
%
%
%

\section{Introduction}
During the last few years, we have seen proposals and experimental implementations of quantum information processors for various  quantum systems \cite{ladd2010}. 
While there is still no ultimate physical platform, nor an underlying information processing model, most of the known candidates must deal with adverse quantum effects. 
Nuclear Magnetic Resonance (NMR) techniques have been successfully used to implement quantum protocols, \eg, quantum teleportation \cite{nielsen1998} and Shor's \cite{vandersypen2001} algorithm. 
NMR remains indispensable in quantum computing as it continues to provide new insights, new methods and new techniques, as well as allowing for testing interesting quantum computing tasks, although there are still huge difficulties in initialization, error-correction and scalability \cite{nielsen2000}.

In the NMR Ensemble Quantum Computing (EQC) introduced by Cory \emph{et al.} \cite{Cory1997}, a key observation  is that, compared to the ideal case where one hundred percent of
the molecules would be polarized,   the proportion of  molecules effectively contributing to
the  magnetisation in a pseudo-pure state is about one part in a million,  nonetheless one still has a macroscopic detectable signal.  On the other hand, when 
performing an algorithm with EQC, one has to deal with two intrinsic limitations of the liquid state NMR, namely: $(i)$ the natural relaxation process the spins are subjected to is
 non-unital, which means that the state simulated by a pseudo-pure state has a  non-linear time evolution, what can compromise the quantum computation.
$(ii)$ Related to this fact, there is  the recovery of the magnetisation towards equilibrium, characterized by the spin-lattice relaxation time $T_1$, which diminishes the amplitude
of the component corresponding to the simulated state   
along the time. This means that expectation values measured in different rounds of the experiment will not be comparable, unless one
takes into account the reduction of their magnitudes along the time. When the quantum computation lasts about  the transverse relaxation time $T_2$, 
the typical coherence time,  these drawbacks are negligible, and one can ignore
both the non-linear evolution and  the degradation of the simulated state, and also some particular expectation value can be taken as a reference in order to normalize the other measurements in the computation.
 Notwithstanding, the spins natural relaxation is an ever present process ($\Phi_{relax}$) superimposed to the map of the particular computation one is performing. 
In the present work, our goal is to characterize the relaxation process in the same way one would reconstruct the map corresponding to a quantum computation,  
\ie, by means of quantum process tomography with the least possible assumptions. We will be able to quantify, along the time, how much the relaxation process departs from being  unital and
trace preserving, and we will also characterize how the relaxation process affects the normalization of expectation values. This way we will learn in which time window it is safe to assume the relaxation process is approximated unital or trace preserving, complementing other previous studies
\cite{Cory2003,Cory2012,Gavini}. Thus we 
hope to shed some light in  how one could compensate for the two drawbacks aforementioned, and consequently  increase the time window for quantum simulations in 
liquid state NMR. The present work also serves the purpose to test,  in a real NMR experiment, and  in a non trivial situation, the efficiency,  robustness, flexibility and simplicity of
implementation of our quantum tomography techniques \cite{maciel2011, maciel2012, maciel2013}.

Robust techniques for quantum process tomography in NMR EQC have been developed in the case the process is assumed to be Markovian and trace preserving \cite{Cory2003} , or when some
model is assumed for the Hamiltonian governing the process \cite{Cory2012}. In both cases, the process tomography was performed in a time window of about  $T_2$, and therefore
the process could be  safely assumed to be unital.  We, on the contrary, will assume the worst case scenario, meaning the process may be non-unital, non-Markovian, and not trace 
preserving. Besides that, we want to follow the quantum process for very long times, much more than the  spin-lattice relaxation time $T_1$,  what  raises the problem of how
to compare measurements taken in different times, which is what we call the \emph{normalization problem}. A pedagogical discussion of why the state simulated by a pseudo-pure state evolves non-lienarly
under the relaxation process, and an \emph{ad hoc} phenomenological  method to circumvent  the \emph{normalization problem} were presented by Gavini-Viana \emph{et al.} \cite{Gavini}. The basic
idea is that a  pseudo-pure state, according to Cory \emph{et al.} \cite{Cory1997}, has the form  
\begin{equation}
\label{pseudo-pure}
 \rho_{pp}=[(1-\alpha) \mathbb{1}_{2^n} + 2\alpha\ketbra{\psi}{\psi}]/[(1-\alpha)2^n+2\alpha], 
 \end{equation}
  where $n$ is 
the number of spin-$\frac{1}{2}$ particles and $\mathbb{1}_{2^n}$ is the $2^n \times 2^n$ identity matrix
 \footnote{One should remember that the pseudo-pure state is not an actual quantum state of the NMR system, but rather the result of  \emph{spatial} or \emph{temporal labelling} 
 \cite{nielsen2000}.}. 
In a quantum computing simulation, we are concerned with  unitary evolutions, which  of course affect only
$\ketbra{\psi}{\psi}$, the simulated state. However,  if $\Phi$ is a non-unital map, $\Phi(\mathbb{1}_{2^n})\neq \mathbb{1}_{2^n}$,  rewriting $\Phi({\rho_{pp}})$ as a pseudo-pure state is equivalent to a non-linear evolution of $\ketbra{\psi}{\psi}$.
Note also that the new pseudo-pure state $\Phi({\rho_{pp}})$, differently from the case of unital evolutions,  has a new ``$\alpha$''. Therefore, the \emph{normalization problem} is equivalent to 
account for the time evolution of ``$\alpha$'' under non-unital processes, such that we can always track the evolution of the resulting pseudo-pure state. 

 To avoid confusion, it is important to remember that a quantum process is a complete positive map, that can be expressed in the usual Kraus sum representation, namely:
 \begin{equation}
 \label{Kraus}
 \Phi\{{\rho}\}=\sum_i K_i \rho K_i^\dagger,
 \end{equation}
 where the $K_i$ are the Kraus operators.  A map is unital if and only if 
 \begin{equation}
 \label{unital}
 \sum_i K_i K_i^\dagger = \mathbb{1}_{2^n},
 \end{equation}
  a condition that does not imply trace preservation. To be trace preserving, 
 a map must satisfiy  
 \begin{equation}
 \label{trace-preserving}
 \sum_i K_i^\dagger K_i = \mathbb{1}_{2^n}. 
 \end{equation}
 The state simulated by a pseudo-pure state evolves linearly under  unital and non-trace preserving maps, 
 and  therefore these maps pose no particular challenge  to usual process tomography
 techniques in liquid state NMR EQC. In particular, the quantum process tomography  method developed by Maciel and Vianna \cite{maciel2012}, 
 and experimentally tested in Quantum Optics by 
 Vianna \emph{et al.} \cite{Roma2013}, could be employed without any modification to handle these maps, as well as non-unital ones. A caveat about our basic
 assumption of the complete positiveness (CP) of the map is that the tomographed map could eventually be non-CP  due to noisy measurements \cite{Cory2004}, 
or in the case of correlations between the initial prepared state and the environment \cite{Ringbauer2015}.

The  methods we will employ here \cite{maciel2012, maciel2011,maciel2013} exploit the natural constraint of positive semidefiniteness arising in quantum tomography, in tandem with semidefinite programs (SDP) \cite{boyd2004}, which are
convex linear optimization problems with linear matrix constraints. A great merit of an SDP is that to find its  global minimum there are many available implemented numerical methods which converge
efficiently in polynomial time \cite{boyd2004, mosek, sedumi98}.  Our tomographic methods \cite{maciel2012, maciel2011,maciel2013} are very easy to implement and are 
also very flexible. They consist of
the minimization of some figure of merit, expressed as a  linear objective function, subject to the constraint of the positive semidefiniteness of the density matrix, in the case of state tomography, 
or the positive semidefiniteness of the dynamical or Choi matrix, in the case of process  tomography, with the restriction that the resulting state or process be compatible
with the measurements. The choice of the objective linear function to be minimized makes our method equivalent to other approaches. For example, by choosing  the nuclear (trace) 
norm
of the positive semidefinite matrix of interest as the objective function, we have a matrix completion approach equivalent to  compressed sensing \cite{gross2010}. Note
however, that nuclear norm minimization is also a good heuristic for quantum tomography in general, including the case of informationally complete measurements.
 Detailed comparison
of our methods with maximum likelihood and maximum entropy was reported in \cite{maciel2013}. Our methods were   tested for state and process tomography 
in Quantum Optics
experiments \cite{Chile2010, UFMG2013, Roma2013}, and in liquid state NMR it was tested for state tomography \cite{oliveira2011} .

The paper is organized as follows. In the next section,  we describe the liquid state NMR setup we use to simulate two qubits, and the basic  theory modelling our system.  Section 3
is dedicated to a discussion of the normalization problem. Section 4 contains our tomographic methodology. In Sec. 5,  we present and discuss  the results. We conclude in Sec. 6.

\section{The NMR system}
The  NMR experiment was performed on a liquid-state enriched carbon-13 Chloroform sample, at room temperature, in a Varian 500 MHz shielded spectrometer.
This sample exhibits two qubits encoded in the $^1$H and $^{13}$C  spin-$\frac{1}{2}$ nuclei.
The heteronuclear interaction of the two spins is characterized by the magnitude of the $J$-coupling being much smaller than the resonant frequencies of the two nuclei, 500 MHz and 125 MHz, for $^1$H and $^{13}$C, respectively.
This system is useful for the simulation of quantum information processing in a Hilbert space of two qubits. 

The quantum model adopted in room temperature liquid-state NMR experiments is of an ensemble of N ($\mathcal{O}(10^{15})$)  non-interacting
independent molecules. Therefore the Hamiltonian can be written as
\begin{eqnarray}\label{N-hamiltionian}
H_{\{N\}} &= H \otimes \mathbb{1}^{\otimes N-1} + \mathbb{1} \otimes H \otimes \mathbb{1}^{\otimes N-2} + \dots \nonumber \\
&\phantom{=} \dots + \mathbb{1}^{\otimes N-2} \otimes H \otimes \mathbb{1} + \mathbb{1}^{\otimes N-1} \otimes H \nonumber \\
H_{\{N\}} &= \sum_{i=1}^N \bigotimes_{j=1}^N \left( \delta_{ij} H + (1-\delta_{ij})\mathbb{1} \right), 
\end{eqnarray} 
where $\mathbb{1}$ is the $2\times 2$ identity matrix,   $\delta_{ij}$ is the Kronecker's delta,  and $H$ is an \emph{effective} Hamiltonian for a single molecule. The wave function and the density matrix corresponding to the N-particle Hamiltonian are assumed to be separable, namely:
\begin{equation}\label{N-state}
\Psi_{\{N\}}=\bigotimes_{i=1}^N \psi_i \quad \mbox{and} \quad \rho_{\{N\}}=\bigotimes_{i=1}^N \rho_i,
\end{equation}
where $\psi_i$ and $\rho_i$ are the wave function and density matrix for a single molecule, respectively, where we shall work only in the spin degrees of freedom. 

In an NMR experiment, one does not have access to the particular state of a distinct molecule in the preparation procedure, all one has is the average state of a representative molecule of 
the ensemble.  Therefore,  we can write:
\begin{equation}\label{N-state-2}
\rho_{\{N\}}=\bigotimes_{i=1}^N \rho_i=\rho^{\otimes N},
\end{equation}
where $\rho$ stands for this \emph{representative state}, which we shall discuss further later.

Our observables are related to the resultant macroscopic magnetisation induced by an external static magnetic field, and we assume they are operators of the form:
\begin{eqnarray}\label{A-N}
A_{\{N\}} &= A \otimes \mathbb{1}^{\otimes N-1} + \mathbb{1} \otimes A \otimes \mathbb{1}^{\otimes N-2} + \dots \nonumber \\
&\phantom{=} \dots + \mathbb{1}^{\otimes N-2} \otimes A \otimes \mathbb{1} + \mathbb{1}^{\otimes N-1} \otimes A, \nonumber \\
A_{\{N\}} &= \sum_{i=1}^N \bigotimes_{j=1}^N \left( \delta_{ij} A + (1-\delta_{ij})\mathbb{1} \right).
\end{eqnarray}
 Therefore, we have expectation values of the kind:
\begin{eqnarray}\label{exp-A-N}
\langle A_{\{N\}} \rangle&= \sum_{i=1}^N \trace{ \rho^{\otimes N} \left[ \bigotimes_{j=1}^N \left( \delta_{ij} A_i + (1-\delta_{ij})\mathbb{1} \right) \right] } \nonumber \\
&= N \, \trace{\rho A}. 
\end{eqnarray}

This rationale, assuming Eqs. (\ref{N-hamiltionian}-\ref{A-N}) as premisses, formalises a model in which one has equally prepared, but independent, black boxes and the same measurement ``buttons'' are pressed in all of them when the measurement occurs. Thus, we have a \emph{reducible} to a single box model and we can generalise and specialise our arguments back and forth, \ie, what happens in a single box, will happen in the other ones and \emph{vice versa}. We shall use this freely.
This model fits the common knowledge that measurements in NMR have contributions of all the molecules in the ensemble \cite{nielsen2000, Cory1997, oliveira2011}, justifying the detected  macroscopic signals. 

The observables of interest in the NMR experiment  are generally the components of the macroscopic nuclear magnetisation, which are proportional to the ensemble average values of the nuclear spin operators. The matrix representation of these operators in the $\ket{S,m}$ basis (with $S=1/2$) reads
$S_x=\sigma_x/2$, $S_y=\sigma_y/2$, $S_z=\sigma_z/2$, with the usual Pauli matrices given by:
\[
 \sigma_x =  \left[ \begin{array}{cc}
0 & 1 \\
1 & 0 \\
\end{array}
\right];
\quad
 \sigma_y =  \left[ \begin{array}{cc}
0 & -i \\
i & \phantom{-}0 \\
\end{array}
\right];
\quad
 \sigma_z =  \left[ \begin{array}{cc}
1 & \phantom{-}0 \\
0 & -1 \\
\end{array}
\right].
\]
For example, one can calculate the magnetisation in the $z$-direction of an ensemble of nuclear spins using the following relation:
\begin{equation}\label{magnetisation_trace_relation}
\avg{M_z} \equiv \hat{I}_z \propto \trace{\rho \sigma_z},
\end{equation}
where $\hat{I}_z$ is the empirical value associated with the observation of the magnetisation in the $z$-direction and $\rho$ is the density matrix of the system of a single specimen. Remember that the left-hand side of this equation represents the statistical average over the entire ensemble of N molecules, and not the expectation value for a given particular member of the ensemble.

The relevant effective two-spin  Hamiltonian in Eq. \refe{N-hamiltionian} is given by
\begin{equation}\label{hamiltonian_NMR}
H = -\hbar \omega_H S_z \otimes \mathbb{1} -\hbar \omega_C \mathbb{1} \otimes S_z +2 \pi \hbar J S_z \otimes S_z,
\end{equation}
where the labels H and C stand for hydrogen and carbon, respectively, as the first and second particles. The resonant frequencies $\omega_H$ and $\omega_C$ include the effects of isotropic chemical shift $\sigma_{iso}$ in acetone solution for each nucleus, \ie, $\omega = \omega_L(1-\sigma_{iso})$. The Larmor frequency  is given by $\omega_L \equiv \gamma_n  B_0$, where $\gamma_n$ is the gyromagnetic ratio of the nucleus,  and  $B_0$ is the static magnetic field in $z$-direction.

At thermal equilibrium, the density matrix obtained from the Hamiltonian in Eq. \refe{hamiltonian_NMR} is given by the maximum entropy principle \cite{jaynes1957a}:
\begin{equation}\label{rho_maxEnt}
\rho_{eq} = \frac{e^{-H/k_B T}}{\trace{e^{-H/k_B T}}},
\end{equation}
where $k_B$ is the Boltzmann constant and T is the temperature. In the high temperature limit, one can expand this exponential as
\begin{equation}\label{rho_eq}
\rho_{eq} \approx \frac{1}{4}\mathbb{1} \otimes \mathbb{1} + \frac{1}{4} \frac{\hbar \omega_H}{k_B T} S_z \otimes \mathbb{1} + \frac{1}{4} \frac{\hbar \omega_C}{k_B T} \mathbb{1} \otimes S_z + \frac{1}{2} \frac{\hbar \pi J}{k_B T} S_z \otimes S_z.
\end{equation}
Note that the last term of Eq. \refe{rho_eq} is usually neglected (see for example \cite{nielsen2000, oliveira2011}),  for it is actually very small compared to the other terms.
Notwithstanding, as we will show, this term contains crucial information about the time window for quantum computing simulations.

Now, we are able to calculate the total magnetisation of the system in the $z$-direction, using the equilibrium state in Eq. \refe{rho_eq}, the relation in Eq. \refe{magnetisation_trace_relation} and Curie's law, which leads to
\begin{equation}\label{total_magnetisation}
M_z^{eq} = \frac{n_H \gamma_H^2 \hbar^2 B_0}{4k_B T} + \frac{n_C \gamma_C^2 \hbar^2 B_0}{4k_B T}  + \frac{n_J \pi^2 \iota_{J}^2 \hbar^2 B_0}{k_B T}, 
\end{equation}
where $n_H$ and $n_C$ are the number of hydrogen and carbon nuclei per unit  volume, respectively, which are equal and correspond to the number
of chloroform molecules per unit volume ($n_H=n_C=n$); $n_J$ stands for the number of \emph{responsive} pairs of $J$-coupled nuclei per unit  volume, and $\iota_{J}$ its ``gyromagnetic ratio''. These responsive pairs of spins correspond to a small fraction of chloroform molecules ($N_J/N \ll 1$), for which the spins of  $^1$H and $^{13}$C are in a chosen coherent state prepared by a certain sequence of radiofrequency pulses (RF). The first issue on normalizing the experimental data arises due to the fact that nor $N_J$, neither its density $n_j$,  are straightforwardly revealed by the magnetisation measurements. We shall discuss this later.



Let us focus on the experiment now. We know that interesting NMR quantum information processing experiments happen in non-equilibrium, where the responsive  $J$-coupled pairs of spins play an important role. In our experiment, we prepare the system in a desired non-equilibrium magnetisation state at time $t=0$ and allow it to relax.  First, at $t=0$, a certain sequence of RF pulses prepares the system  in some desired non-equilibrium state,  where we have the maximum number of responsive $J$-coupled spin pairs.
 This results in a small macroscopic non-equilibrium magnetisation. 
 The second part of the experiment happens in the time window which we can use the small fraction of responsive molecules in the non-equilibrium state to simulate 
the dynamics of quantum parameters (\eg, \cite{vandersypen2001}). The system is allowed to relax, the molecules that responded to
the preparation start to recover their equilibrium state, diminishing $N_J$, and finally ($t \rightarrow \infty$ ) the system reaches the equilibrium with vanishing 
$N_J$.

The \emph{relaxation} is such that, after some time has elapsed, the magnetisation returns almost completely to the initial $z$-direction, satisfying again the thermal equilibrium requirements. It is worth emphasising that the exact meaning of ``some time'' is largely dependent on the details of each particular nuclear spin system and its environment, ranging typically from microseconds to several hours. The relaxation consists of two different processes, occurring simultaneously but, in general, independently,  namely: the \emph{transverse relaxation} and the \emph{longitudinal relaxation} \cite{oliveira2011}. 
We shall characterize, by the quantum process tomography, a net effect of these two
 phenomena, that is paramount to the problem of NMR simulation
of quantum information processing.  For reasons to become clear soon, we shall refer to this net phenomenon as the \emph{J-coupling relaxation}.

The longitudinal  relaxation is the process that leads the longitudinal component of the nuclear spin magnetisation to recover its equilibrium value.  The recovery of the $M_z$ component of the magnetisation is related to transitions between the nuclear spin levels. The natural tendency is the system to give up its excess of energy by effecting transitions  from the upper to the lower energy level. After some time, which is commonly named $T_1$, the Boltzmann distribution is almost reestablished.

Simultaneously to the longitudinal relaxation, the transverse relaxation is the process that leads to the disappearance of the components of the nuclear magnetisation $\mathbf{M}$ that are perpendicular to the static field $\mathbf{B}_0$. The origin of the transverse relaxation relies on the loss of coherence in the precession motion of the spins (or dephasing of the spins), caused by the existence of spread in precession frequencies for the collection of nuclear spins. This spread progressively results in a reduction of the resultant transverse components (\eg, $M_x$ and $M_y$). After some time, which is commonly named $T_2$, the spins distribute randomly in a precession cone around $\mathbf{B}_0$ and the transverse magnetisation tends to  zero.

The $J$-coupling relaxation, which we introduce here, can be considered as the net effect of both longitudinal and transverse relaxations.
 In NMR quantum information processing simulation experiments, correlated spins in non-equilibrium states are indispensable  to \emph{simulate} quantum algorithms. We name $T_1^J$ and $T_2^J$ the characteristic times in which the coupled magnetisations $M_{ij} \, (\forall \, i,j \in \{ x,y,z \} $) decay,  $T_1^J$ for the longitudinal, or $M_{zz}$ component,  and $T_2^J$ for the transverse components.  Note that what we call \emph{coupled-magnetisation} refers
specifically to expectation values of two-spin operators $\left( \trace{\rho \sigma_i^H \otimes \sigma_j^C} \right)$, while \emph{uncoupled-magnetisation} is related to expectation values of
one-spin operators $\left( \trace{\rho \sigma_i^H \otimes\mathbb{1}} \mbox{ or } \trace{\rho \mathbb{1}\otimes \sigma_j^C} \right)$.  
Our motivation for introducing $T_1^J$ and $T_2^J$ was the following.  (\emph{i}) These parameters can be extracted from the usual NMR data;  (\emph{ii}) the $M_{zz}$ component behaves distinctly from all the other coupled magnetisation components; (\emph{iii}) and finally, these parameters   characterize an important
 change in the behaviour of the relaxation process map, as we will show.
  
In order to reconstruct the density matrix of the system along the time, we need to normalize the experimental data, such that we can relate the measured 
  non-equilibrium magnetisations to expectation values. Here, we are not concerned with  the simulated  states, we deal only with the
  real  non-equilibrium NMR states, although it is worth to emphasise that, in the preparation procedure, we also assume \emph{ergodicity}, as we collect the whole set of measurements using \emph{temporal labeling} \cite{nielsen2000,oliveira2011,leskowitz2004}.

 \section{The normalization problem}
 
 The normalization issue is a daunting problem in NMR experiments. Firstly, by the data acquisition itself, since it  involves taking the intensity of the peaks from a Fourier transform spectra of an oscillating signal. Secondly, because we are forced to assume the representative state $\rho$ in Eq. \refe{N-state-2}. It would be preferable if we could address each molecule individually like
\[
\rho_{\{ N \}}(t) = \rho_{1}(t) \otimes \rho_{2}(t) \ldots \otimes \rho_{N}(t)  .
\]
If it were the case, we would be able to count how many spins are still coherent and responding accordingly, therefore it would be possible to know exactly the number of responding molecules as in Eq. \refe{exp-A-N} (\eg, if we were interested in knowing $N_J$). However  such a case is not possible and we shall stick to the weaker premise of the representative state, losing track of this counting. Thirdly, as consequence of the traceless nature of the operators corresponding to the magnetisation components, we do not possess any direct experimentally measured  data related to the trace of the non-equilibrium density matrix. If we were measuring the eigenprojectors  
($P_k$) of these operators ($A=\sum_{k=1}^d \lambda_k P_k$), this problem would not appear, as the completeness relation ($\mathbb{1}_d=\sum_k P_k$) would give us a trivial recipe for normalization, \ie, we would impose $\sum_{k=1}^d \trace{\rho P_K}=  1$. However, what we have is $\mean{A}=\trace{\rho A}$, and for this measurement to be useful, we need
some reference expectation value $\mean{A_{ref}}$. Of course, the reference expectation value cannot vary in time. For example, in the EQC MMR experiment reported in \cite{oliveira2011},
which employed the same experimental setup and molecule of the present work, the reference expectation value was   $\mean{\sigma_Z^H \otimes \sigma_Z^C}$ in a pseudo-pure state simulating
the two-qubit state $\ket{\uparrow \uparrow}$. It was imposed then that $\mean{\sigma_Z^H \otimes \sigma_Z^C} / N_{ref}= 1$, and all other expectation values were divided by  $N_{ref}$. 
  As we shall see, the quantum tomography techniques we employ here \cite{maciel2011, maciel2012, maciel2013}, allow us to circumvent this problem, yielding proper inferences of the normalized non-equilibrium quantum
  states and quantum maps. 

\begin{figure}[!ht]
\centering
\includegraphics[width=7cm]{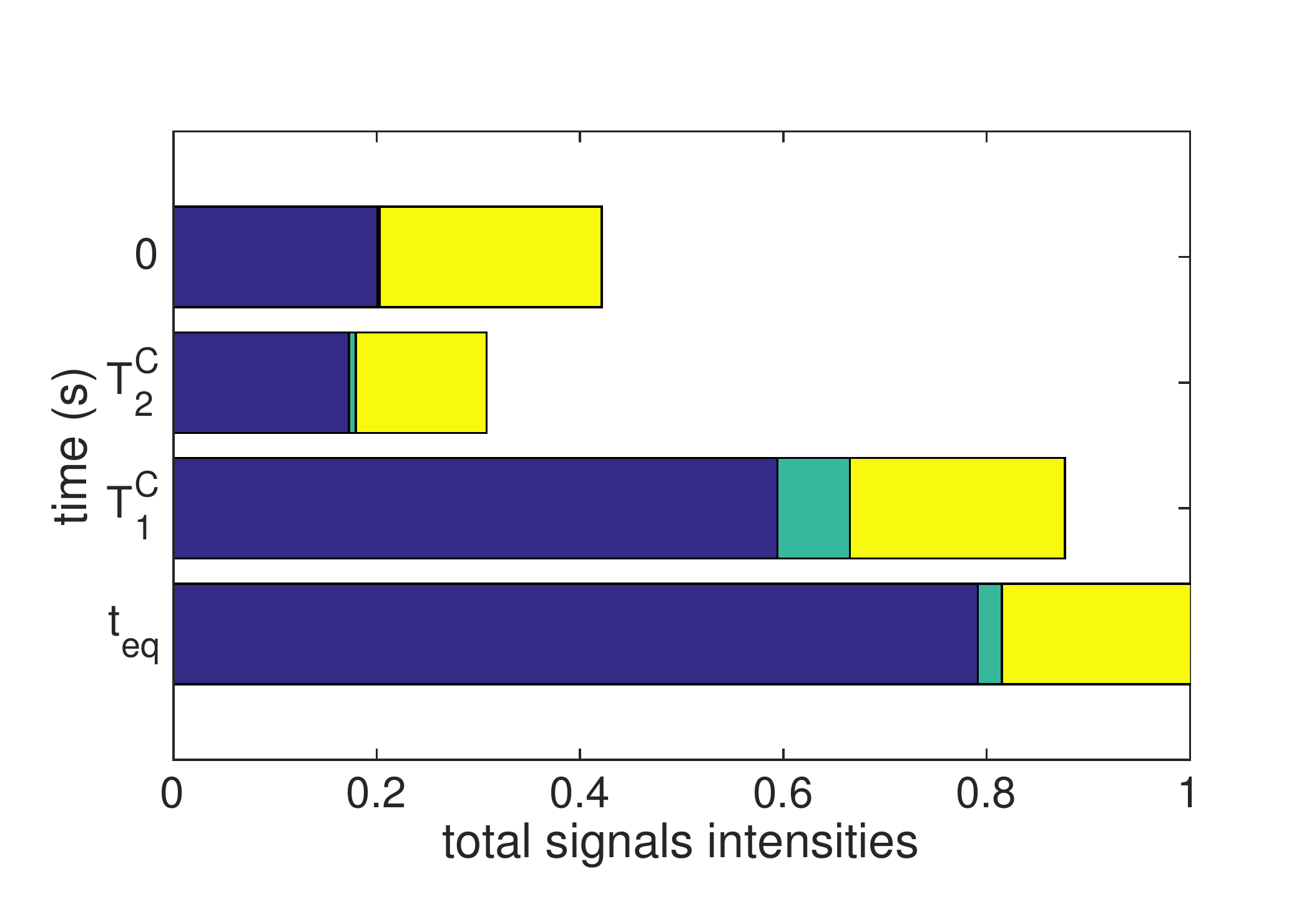}
\caption{Comparison of the three distinct total signals intensities along time in the experiment. (Blue) Total hydrogen signals: 
$I^H= \sum_{i=x,y,z}|\hat{I}_i^H(t)|$ (\cf Eq. \refe{normalize}); (Yellow) total
carbon signals: $I^C=\sum_{i=x,y,z}|\hat{I}_i^C(t)|$ (\cf Eq. \refe{normalize});
 (Green) total coupled signals: $I^J=\sum_{i,j=x,y,z}|\hat{I}_{ij}^J(t)|$ (\cf Eq. \refe{normalize-2}).} 
\label{population_diagram}
\end{figure}

We have three different kinds of signals corresponding to our measurements, which we have to deal with in order to normalize the experimental data. One may grasp this in Fig. \ref{population_diagram}.  We have, responding to RF pulses in the resonant frequency $\omega_H$ of the hydrogen, both coherently coupled and  uncoherently coupled hydrogen nuclei; the same happens to carbon, for RF pulses in  the resonant frequency  $\omega_C$. As the gyromagnetic ratio of the hydrogen is almost $4$ times larger  than that of the carbon, the hydrogen species is expected to have a larger magnetisation in equilibrium. The last kind of signal, which is very small compared to the first two, corresponds to that coherently coupled magnetisation of the $N_j$ chloroform molecules.
To remedy the lack of some relation tying the intensity of measured magnetisation components, 
like the completness  relation we have mentioned in the previous paragraph, 
we normalize all measured signals \cite{leskowitz2004} corresponding to non-equilibrium magnetisations backwards, such that the expected value of the magnetisation in equilibrium is equal to one.  Therefore, the equilibrium magnetisation plays the role of our reference expectation value that does not vary in time, as discussed in the previous paragraph. This rescaling of the magnetisation signals  is  also convenient in order to perform our numerical optimizations with good precision. 
For the hydrogen and carbon signals, we have
\begin{equation}\label{normalize}
\hat{I}_i^H(t) = \frac{1}{M}\frac{\hbar \omega_H}{4k_bT}\frac{\hat{I}_i^{H_{exp}}(t)}{\hat{I}_{z}^{H_{eq}}} \, \mbox{and} \, \hat{I}_j^C(t) = \frac{1}{M} \frac{\hbar \omega_C}{4k_bT}\frac{\hat{I}_j^{C_{exp}}(t)}{\hat{I}_{z}^{C_{eq}}},
\end{equation}
and for the coupled signals we have
\begin{equation}
\label{normalize-2}
\hat{I}_{ij}^J(t) = \frac{1}{M}\frac{\hbar \pi J}{2k_bT}\frac{\hat{I}_{ij}^{J_{exp}}(t)}{\hat{I}_{zz}^{J_{eq}}},
\end{equation}
where
\[
M = \frac{\hbar(\omega_H + \omega_C +2 \pi J)}{4k_b T},
\] 
and $\hat{I}_{i,j}^{\{H,C,J\}_{exp}}(t)$ are the empirical values associated with the observation of the magnetisation in the $ i,j \in \{ x,y,z \} $ directions relative to the hydrogen, carbon and J-coupled signals respectively. The time evolution of these three kinds of signals   is presented  schematically in Fig. \ref{population_diagram}. In the figure one can see that
at $t=0$, the hydrogen and carbon have the same magnetisation, as a consequence of the state preparation; at $t=T_2^C$, the coherence has almost vanished; at $t=T_1^C$, the 
system has a magnetisation reminiscent of the equilibrium, with the total hydrogen signal greater than that of the carbon; and at $t=60s$, which we call $t_{eq}$, the system is practically thermalized, with the total signal equal to one, and the hydrogen signal almost 4 times larger than that of the carbon.

The equilibrium (Eq. \refe{rho_eq}) and non equilibrium states dealt with in the NMR experiment are of the kind 
\begin{equation}\label{deviation}
\rho\approx \frac{1}{4}\mathbb{1}\otimes \mathbb{1} + \epsilon \Delta,
\end{equation}
where $\epsilon\approx 10^{-5}$ and $\Delta$ is a traceless matrix, known as the deviation matrix, corresponding to the last three terms of Eq. \refe{rho_eq}. 
Though the experimental magnetisation signals come just from the deviation matrix, due to the traceless nature of the spin operators (for example: $\trace{\rho \sigma_i \otimes \sigma_j}= \epsilon \trace{\Delta \sigma_i \otimes \sigma_j}$), these are macroscopic signals, for there are still  a huge number ($\mathcal{O}(\epsilon \times 10^{23})=\mathcal{O}(10^{18})$) 
of  molecules contributing to the magnetisation.

As discussed in the Introduction,  the time evolution of the state simulated by a pseudo-pure state  (Eq. \refe{pseudo-pure}) is non-linear, if the process is non-unital.  It is easy
to circumvent this non-linearity. We have simply to tomograph the process acting directly on the NMR state given by Eq. \refe{deviation}, which of course evolves linearly.

In summary, our strategy to keep measurements taken in different times  properly normalized and comparable consists of two premisses, namely: (\emph{i}) the quantum tomography
accounts for the process acting on the pseudo-pure state along the time, rather than the process acting on the simulated state; (\emph{ii}) the process at any finite time must be compatible
with the magnetisation converging to unity in an infinite time, \ie, the equilibrium magnetisation is taken as a reference to normalize all other measurements.

\section{The quantum tomographies}
The \emph{standard quantum process tomography}  (SQPT) \cite{chuang1997,nielsen2000} consists in obtaining information about the map  by means of quantum tomography of  states modified by  the action  of the unknown process map. If the density matrices of the tomographed states have dimension $d$,  then we need $d^2$ linear independent
states to reconstruct the quantum  map.  As we are dealing with two qubits, $d$ is 4. Our experiment consists in preparing  20 different 
time average pseudo-pure states \cite{nielsen2000}
  ($\varrho_k^l, \,l=1,2,\ldots, 20$), which  can be used to simulate 20 different two-qubit pure states,  where the density matrices of 16 of them are linear independent, spanning the Hilbert-Schmidt space. We also have 51 observations of these states in the time instants $t_k$ $(k=0, 1, \ldots , 50)$. The simulated 20 pure states correspond to 
  16 states forming the  two-qubit Mutually Unbiased Basis  (MUB) described in  \cite{maciel2009}, plus the four canonical states 
   $\ket{\uparrow \uparrow}$, $\ket{\uparrow \downarrow}$, $\ket{\downarrow \uparrow}$, $\ket{\downarrow \downarrow}$ .
   
{The overcomplete set of 20 density matrices were preferred in order to increase the quality of the process tomography, following the arguments in \cite{burgh2008,langford2013}. 
This strategy of enlarging the number of measurements with an overcomplete basis,  in order to increase precision,  was also used by one of us in the process tomography
of a two-qubit map in Quantum Optics \cite{Roma2013}. 

Since we have a large ensemble and  sharp peaked measured signals $\hat{I}_{i,j}^{\{H,C,J\}_{exp}}(t)$, we can consider the latter  gaussianly distributed with standard deviation $\sigma^2 =1$. Thus, we can deal with the experimental errors with a simple $\ell_2$-norm minimization, which is equivalent to a \emph{log-likelihood} method assuming gaussian noise  \cite{boyd2004, koenraad2009}. 

The first thing to do is to perform a quantum state tomography  in order to have a preliminary estimate of  the initial states. We use a variation of \cite{maciel2011,maciel2013} also equipped with nuclear norm minimization \cite{gross2010} as follows:
\begin{eqnarray}\label{vqtgauss}
\min_{\rho_0^l, \ \Delta} \ \ \ & \norm{\rho_0^l}_* + \sum_{l,i,j} \Delta_{i,j,ij}^{l,\, \{ H,C,J \}} \nonumber \\
\mbox{subject to} \ \ \ & \ \ \, \rho_0^l \ \succeq 0 \,  & \nonumber \\
& \norm{\trace{{\rho}_0^l \, \sigma_i^H\otimes \mathbb{1}}  \,\,\,- \hat{I}_i^{l,\,H}(0)}_2 \le \Delta_i^{l,\,H} \, & \nonumber  \\
& \norm{\trace{{\rho}_0^l \,\mathbb{1} \otimes\sigma_j^C} \,\,\,\,- \hat{I}_j^{l,\,C}(0)}_2 \le \Delta_j^{l,\,C} \, &   \\
& \norm{\trace{{\rho}_0^l \,\sigma_i^H\otimes \sigma_j^C} - \hat{I}_{ij}^{l,\,J}(0)}_2 \le \Delta_{ij}^{l,\,J} \, & \nonumber  \\
& \ \ \,  \forall \,  i,j \in \{ x,y,z \} & \nonumber \\
& \ \ \, \Delta_i^{l,\,H}, \Delta_j^{l,\,C}, \Delta_{ij}^{l,\,J}  \geq 0 \, . & \nonumber
\end{eqnarray}
where  $\norm{\cdot}_*$ stands for the \emph{nuclear norm} or \emph{trace norm}. This type of \emph{Semidefinite Problem} (SDP) in the second-order cone can be efficiently solved using well known algorithms \cite{yalmip04,mosek,sedumi98}.  

The program in  Eq. \refe{vqtgauss}  has the following interpretation. We want a point estimate ($\rho_0^l$) of a density matrix (positive semidefinite constraint), with minimum residuals 
 ($\Delta_{i,j,ij}^{l,\, \{ H,C,J \}}$)   in the measured data, according to $\ell_2$-norm minimization. 
The normalization of the states \{$\rho_0^l$\} is unknown, and that is why it is convenient to adopt the trace norm minimization. The traces of the states
\{$\rho_0^l$\} are constrained by the normalization we imposed on  the data, \cf Eq. \refe{normalize}  and Eq. \refe{normalize-2}, which corresponds to a unity magnetisation in 
equilibrium, \cf Eq. \refe{total_magnetisation}. 
 
 After this preliminary estimation, we define a lower bound ($N_0$) for the trace of the states, namely:
  \begin{equation}\label{Ntk}
N_0 = max \left\{ \trace{{\rho}_0^l}\right\},   \ l=1,\dots,20.   
\end{equation}
Then we rerun the program in Eq. \refe{vqtgauss} with the additional constraint: $\trace{\rho_0^l}\geq N_0$.
Note that the trace  is related to the population in each state. The largest trace yielded by the first run of the state tomography identifies the population of \emph{responsive molecules}. Therefore $N_0$ is our \emph{reference population}, and we are  interested, at the initial time, in states with the same populations, in order to compare, along the time,
 the different expectation values.

Now we consider the action of the unknown map $\Phi(t_k)$ upon the initial states determined in the second run of Eq. \refe{vqtgauss}, namely: 
\begin{equation}\label{cpmap}
\tilde{\rho}_k^l \equiv \Phi(t_k)\{\rho_0^l\} = \sum_{i,j=1}^{d^2}\chi_{ij}(t_k) A_i \rho_0^l A_j^{\dagger},
\end{equation}
where  the \{$A_m$\} is a complete set of operators forming a basis in the Hilbert-Schmidt space. We obtain the basis $\{A_m\}$ by forming the linearly independent projectors corresponding to the  states in our  two-qubit MUB \cite{maciel2009}.
The \{$\chi_{ij}(t_k)$\} are the elements of a positive semidefinite matrix, and form a representation of the  quantum map in the basis \{$A_m$\} \cite{Zyczkowski}.
 
 We form the Liouville representation of the quantum state by putting the matrix elements of its $d$-dimensional density matrix, in lexicographical order, into a $d^2$-dimensional vector
  $\ket{\rho_0^l}$. We say $\ket{\rho_0^l}$ is a ``\emph{reshape}'' of $\rho_0^l$. The Liouville representation of the map $\Phi(t_k)$, in the operator basis \{$A_m$\},  is a $d^2$-dimensional matrix, namely
  \cite{Zyczkowski}:
  \begin{equation}
  \label{Phi-matrix}
\Phi_k=\sum_{ij}\chi_{ij}(t_k)A_i \otimes A_J^*. 
  \end{equation} 
  Now the action of the map over the state is given by a simple matrix multiplication, namely:
  \begin{equation}
  \label{Phi-rho}
  \ket{\tilde{\rho}_k^l}=\Phi_k\ket{\rho_0^l}.
  \end{equation}
  The density matrix ${\tilde{\rho}_k^l}$ is recovered by the reverse \emph{reshape} operation.
  Now consider $\{\ket{\mu}\}$ and $\{\ket{m}\}$ as two orthonormal basis for a $d$-dimensional Hilbert space. If the matrix elements of $\Phi_k$ are given by:
  \begin{equation}
 (\Phi_k)^{m \mu}_{n \nu}=\bra{m \mu}\Phi_k\ket{n \nu},
  \end{equation}
  then the Choi matrix (or dynamical matrix) $D_k$ of the map is given by a ``\emph{reshuffle}'' of $\Phi_k$, \ie,
  \begin{equation}
  \label{D-matrix}
  (D_k)^{m n}_{\mu \nu} = (\Phi_k)^{m \mu}_{n \nu}.
    \end{equation}
  For a map to be complete positive, its  dynamical matrix must be positive semidefinite ($D_k\succeq 0$). A map is trace preserving if and only if
  \begin{equation}
  \label{trace-preserving-2}
  \sum_{m \mu} (D_k)^{m n}_{\mu \nu} = \mathbb{1}_d, 
  \end{equation}
  and it is unital if and only if
  \begin{equation}
  \label{unital-2}
  \sum_{n \nu} (D_k)^{m n}_{\mu \nu} = \mathbb{1}_d.
  \end{equation}

At each time $t_k$, the quantum process tomography is obtained using a variation of the method in \cite{maciel2012}, also equipped with a matrix recovery technique like nuclear norm minimization,  in order to get the information about the trace:
\begin{eqnarray}\label{process_tomos}
\min_{{\chi(t_k)}, \ \Delta} \ \ \ & \norm{D_k}_* + \sum_{l,i,j} \Delta_{i,j,ij}^{l,\, \{ H,C,J \}} \nonumber \\
\mbox{subject to} \ \ \ &  \chi(t_k) \ \succeq 0 \,  & \nonumber \\ 
&  D_k \succeq 0 \, & \nonumber \\
& \norm{\trace{\tilde{\rho}_k^l \, \sigma_i^H\otimes \mathbb{1}}  \,\,\,- \hat{I}_i^{l,\,H}(t_k)}_2 \le \Delta_i^{l,\,H} \, & \nonumber  \\
& \norm{\trace{\tilde{\rho}_k^l \,\mathbb{1} \otimes\sigma_j^C} \,\,\,\,- \hat{I}_j^{l,\,C}(t_k)}_2 \le \Delta_j^{l,\,C} \, &   \\
& \norm{\trace{\tilde{\rho}_k^l \,\sigma_i^H\otimes \sigma_j^C} - \hat{I}_{ij}^{l,\,J}(t_k)}_2 \le \Delta_{ij}^{l,\,J} \, & \nonumber  \\
& \ \ \,  \forall \,  i,j \in \{ x,y,z \} & \nonumber \\
& \ \ \,  \forall \,  l = 1,\dots,20 & \nonumber \\
& \ \ \, \Delta_i^{l,\,H}, \Delta_j^{l,\,C}, \Delta_{ij}^{l,\,J}  \geq 0 \, . & \nonumber
\end{eqnarray}
The program in Eq. \refe{process_tomos} can be interpreted as follows. We want a minimum trace positive semidefinite estimate of the dynamical matrix ($D_k$), compatible
with the data and having minimal residuals, analogously to the program of Eq. \refe{vqtgauss}. In principle, it is redundant to include the constraint of positive semidefiniteness for both $D_k$ and $\chi(t_k)$, but, in practice, it gave us slightly better numerical precision using the interior-point based solvers MOSEK and SeDuMi \cite{mosek, sedumi98}.

\section{Results and Discussion}
\begin{figure}[!ht]
\centering
\includegraphics[width=9cm]{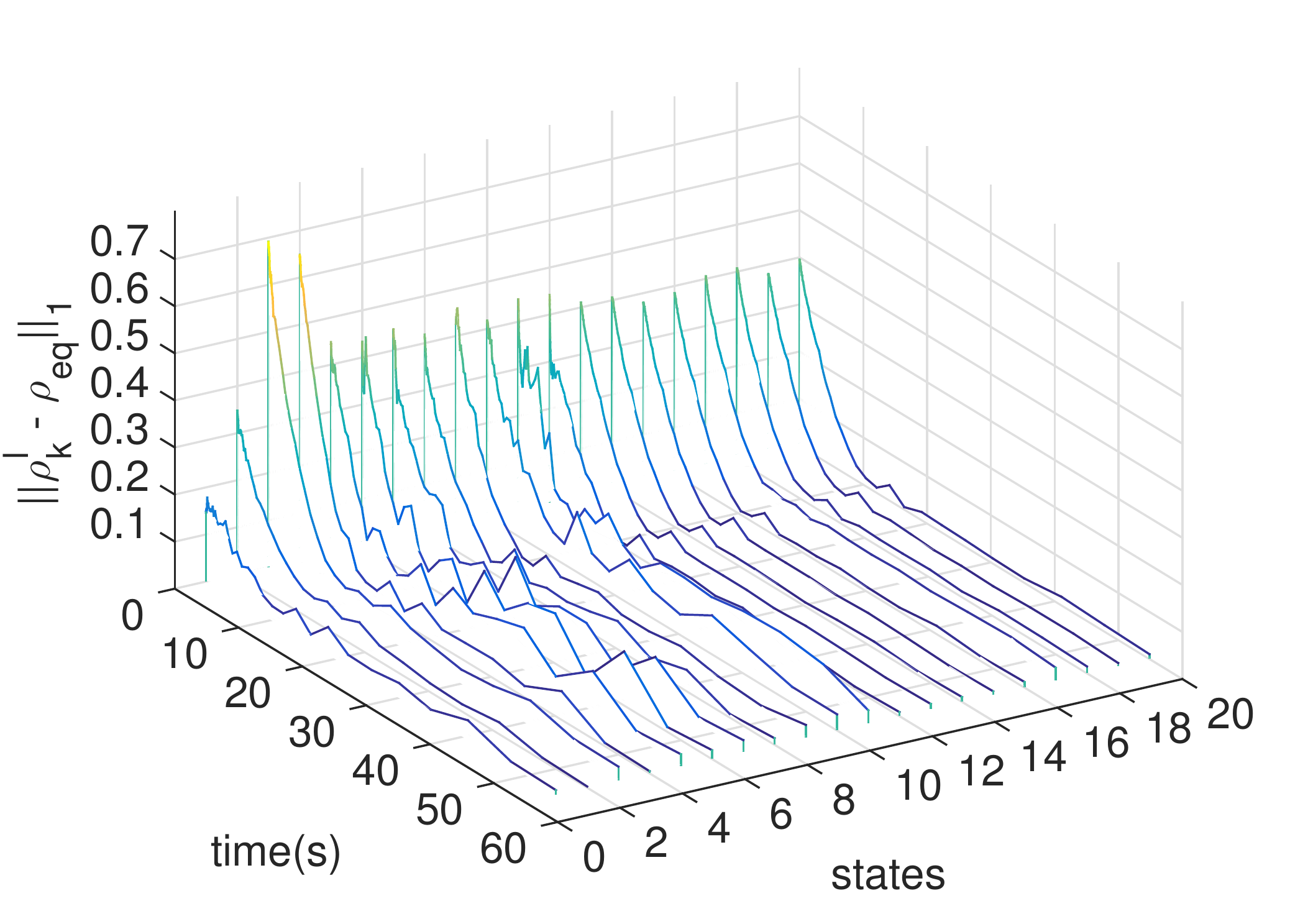}
\caption{The relaxation of the states toward equilibrium. The figure shows the trace distance between the states ${\rho}_0^l$ under the action of the tomographed quantum map $\Lambda(t_k)$ and the equilibrium thermal state. Note that the states in $t=0$ are perturbations of the thermal state by different sequences of radiofrequency pulses. These perturbed thermal states
are the ones useful for quantum information processing simulations. }
\label{figs_relax}
\end{figure}

\begin{figure}[!ht]
\centering
\includegraphics[width=9cm]{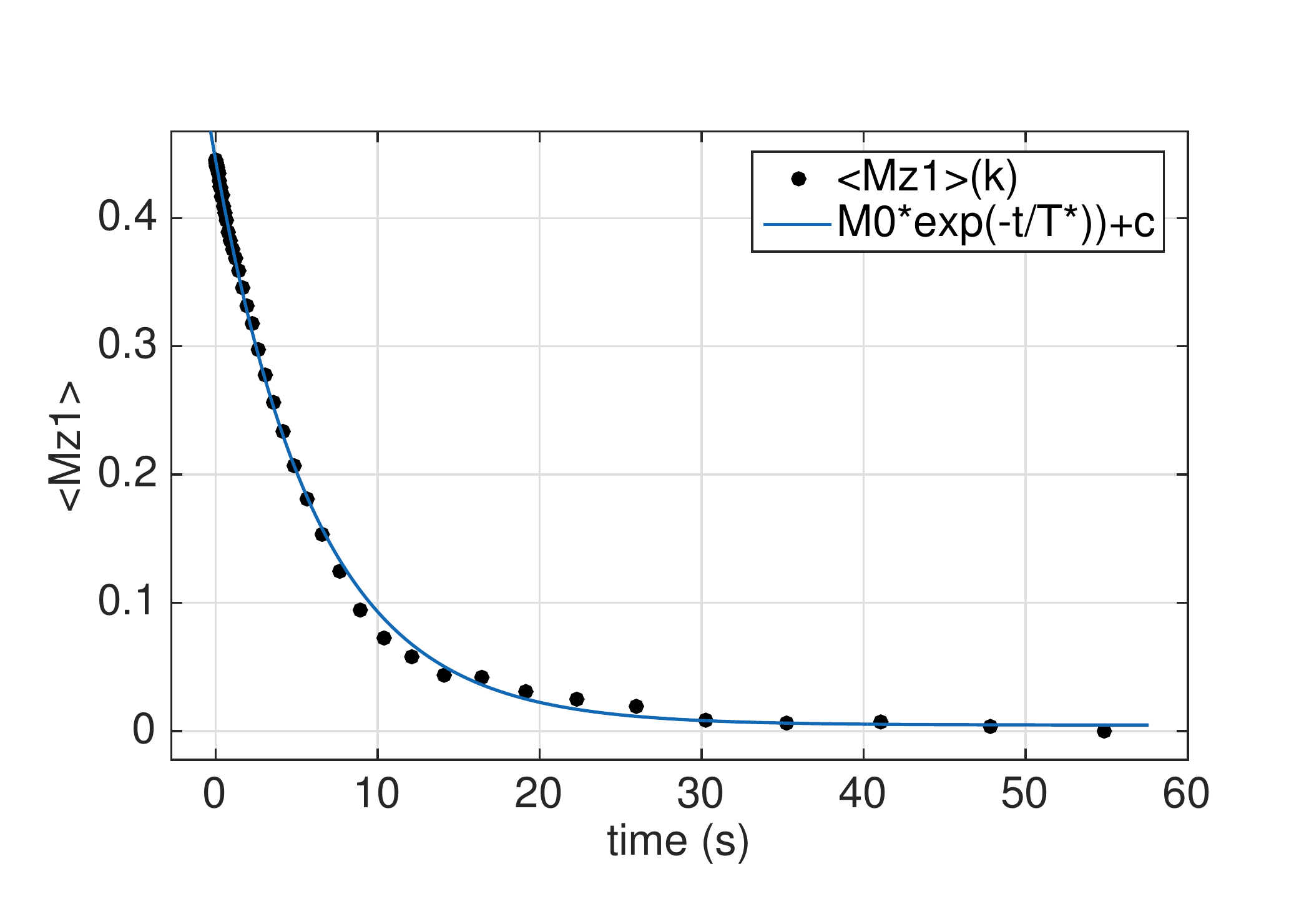}\\
\includegraphics[width=9cm]{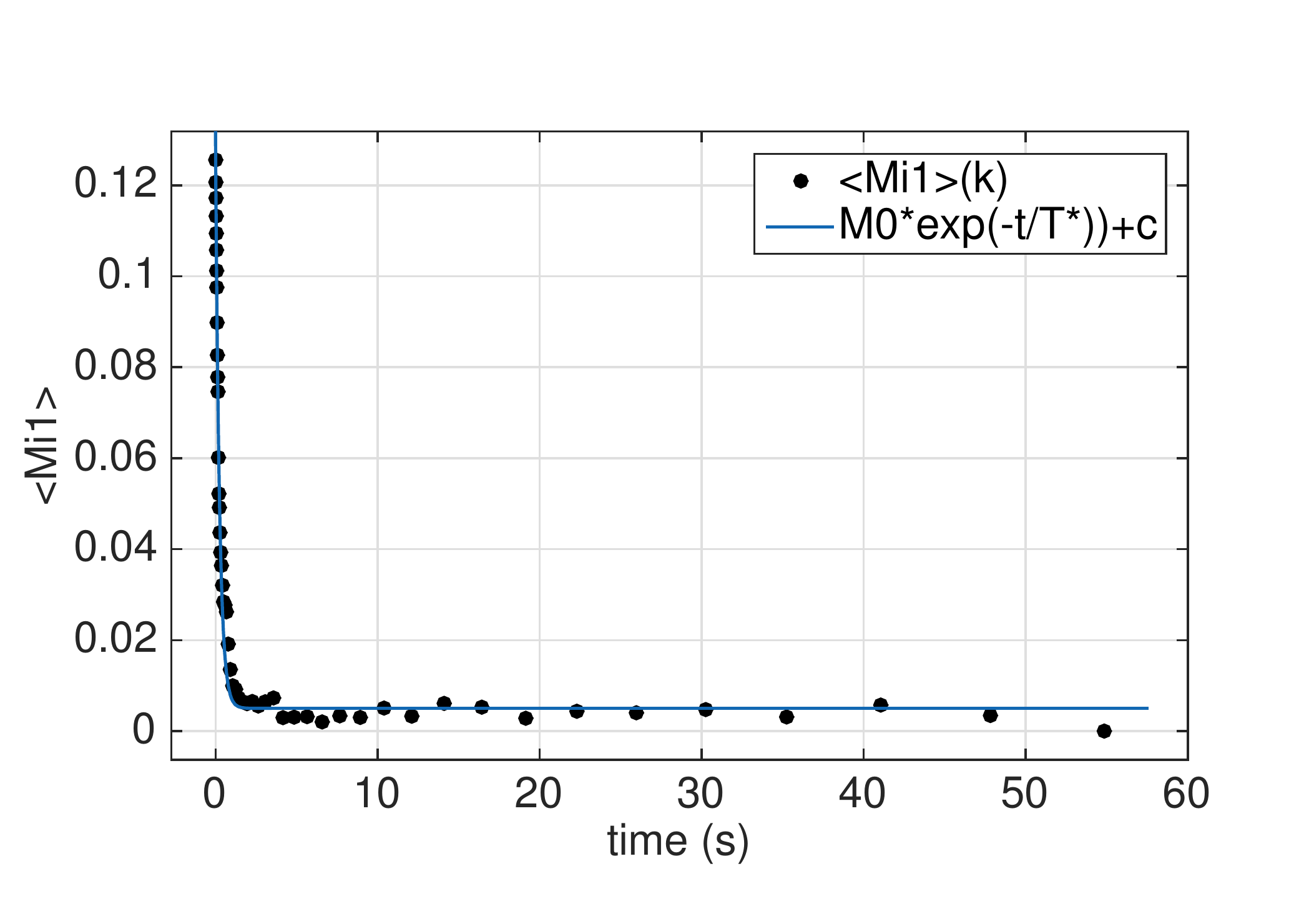}
\caption{The Hydrogen (\emph{uncopled-})relaxation. (\emph{Top}): the quantity $\avg{\tilde{M}_{z\mathbb{1}}}$ as a function of time. (\emph{Bottom}): the quantity $\frac{\avg{\tilde{M}_{x\mathbb{1}}}+\avg{\tilde{M}_{y\mathbb{1}}}}{2}$ as function of time. Note that these ``magnetisations'' are relative to the equilibrium magnetisation, according to Eq. \refe{def_Mj}.}
\label{figs_MH}
\end{figure}

\begin{figure}[!ht]
\centering
\includegraphics[width=9cm]{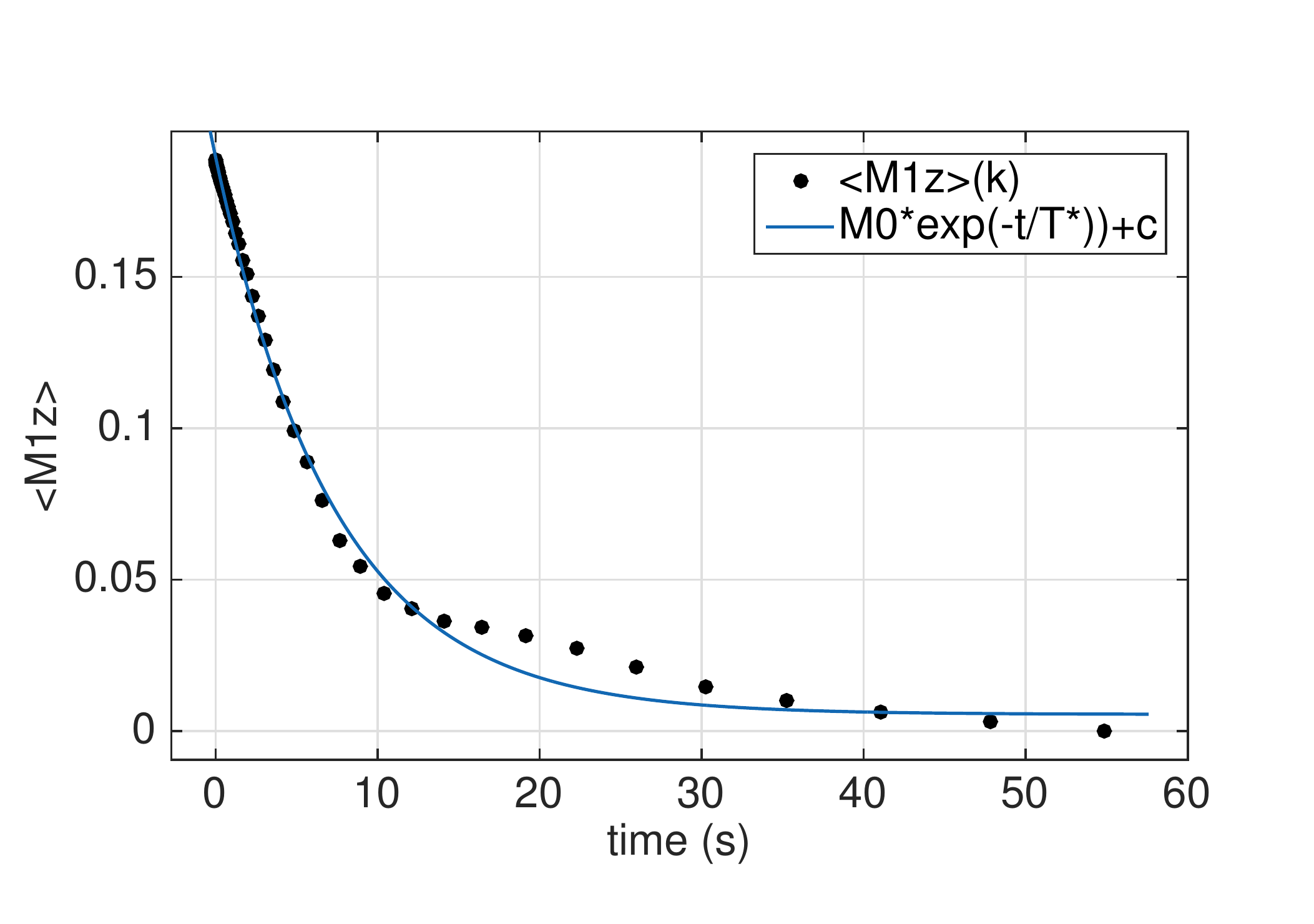}\\
\includegraphics[width=9cm]{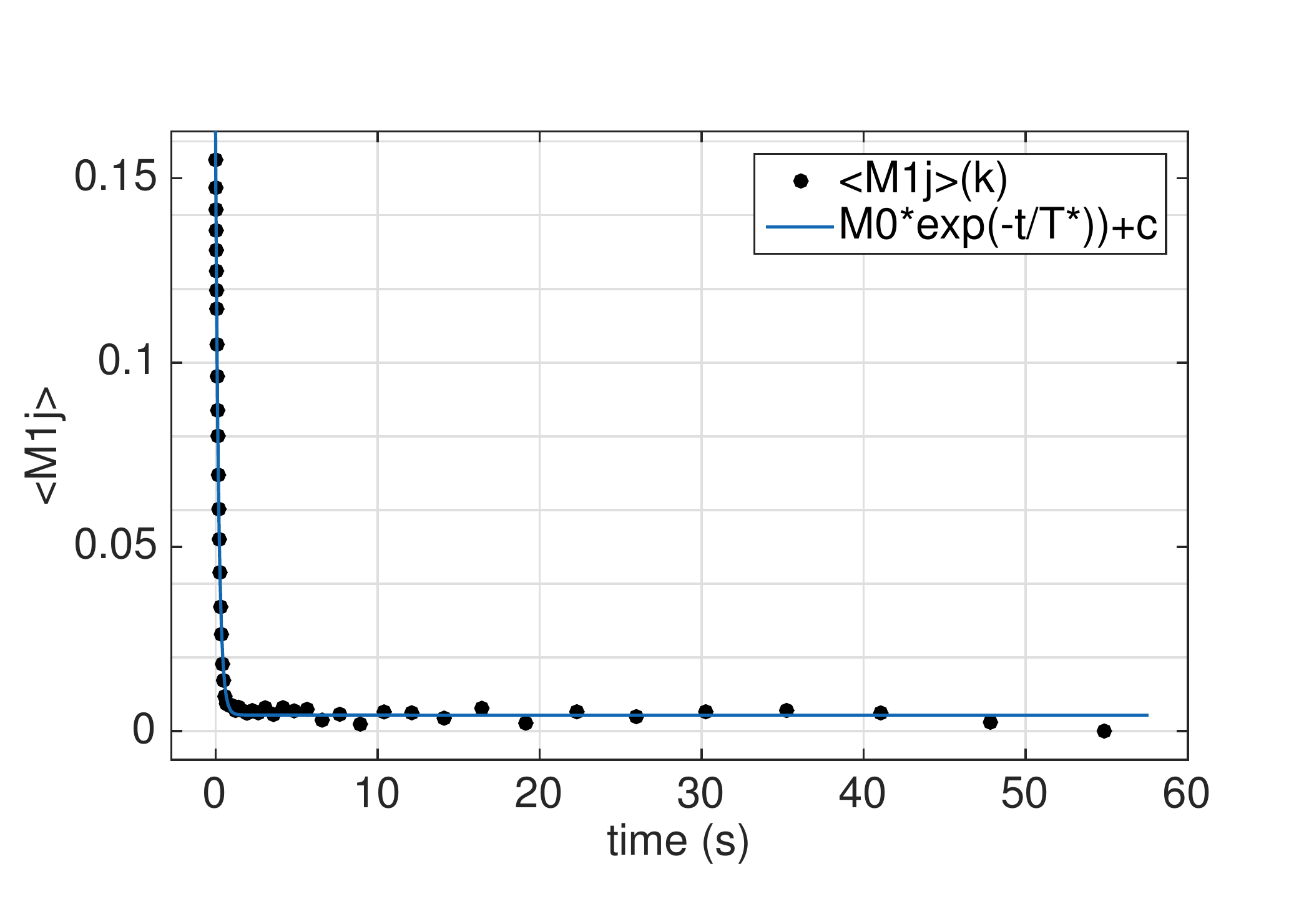}
\caption{The Carbon (\emph{uncoupled-})relaxation. (\emph{Top}): the quantity $\avg{\tilde{M}_{\mathbb{1}z}}$ as a function of time. (\emph{Bottom}): the quantity $\frac{\avg{\tilde{M}_{\mathbb{1}x}}+\avg{\tilde{M}_{\mathbb{1}y}}}{2}$ as a function of time. Note that these ``magnetisations'' are relative to the equilibrium magnetisation, according to Eq. \refe{def_Mj}.}
\label{figs_MC}
\end{figure}

\begin{figure}[!ht]
\centering
\includegraphics[width=9cm]{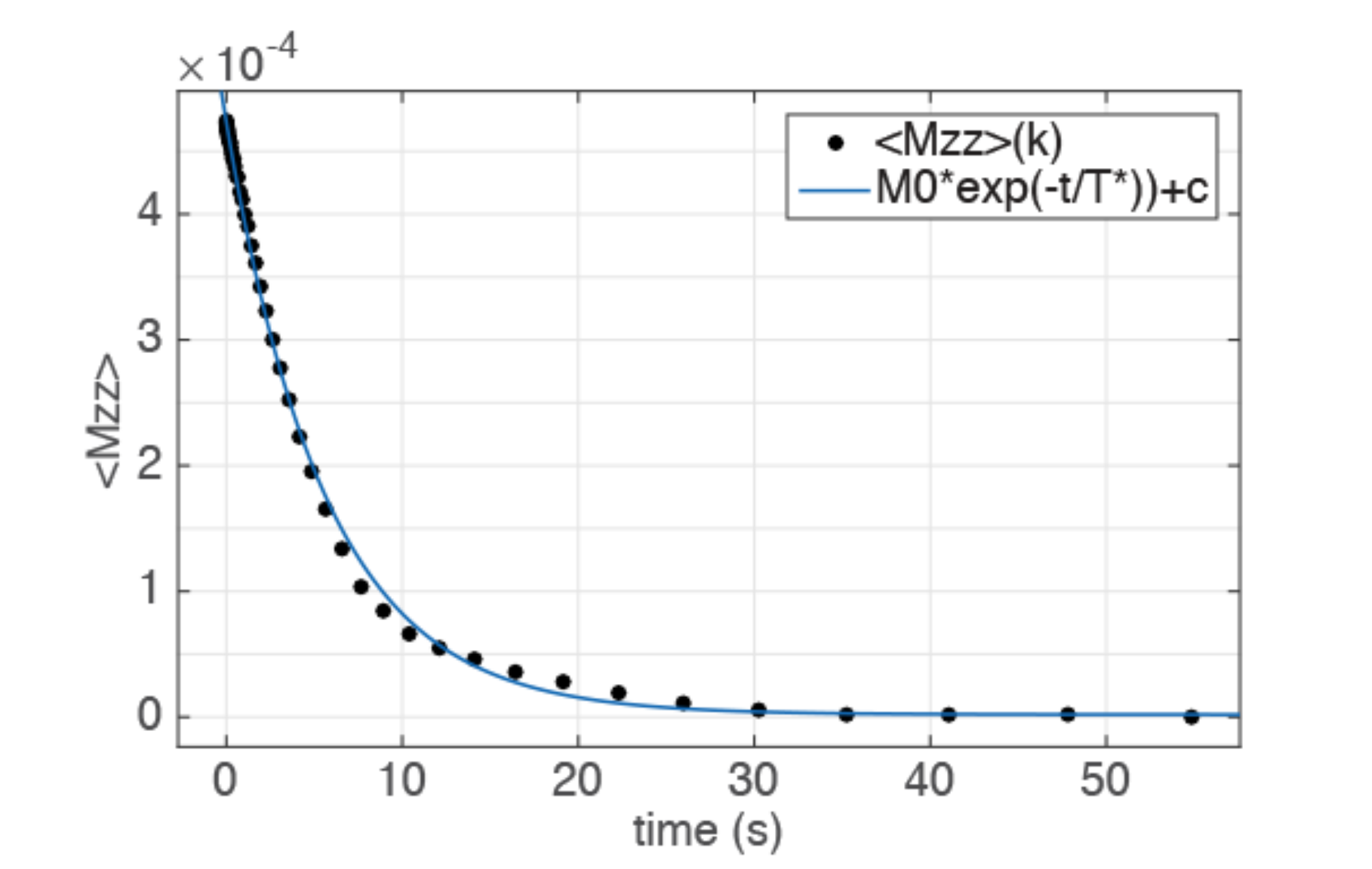}\\
\includegraphics[width=9cm]{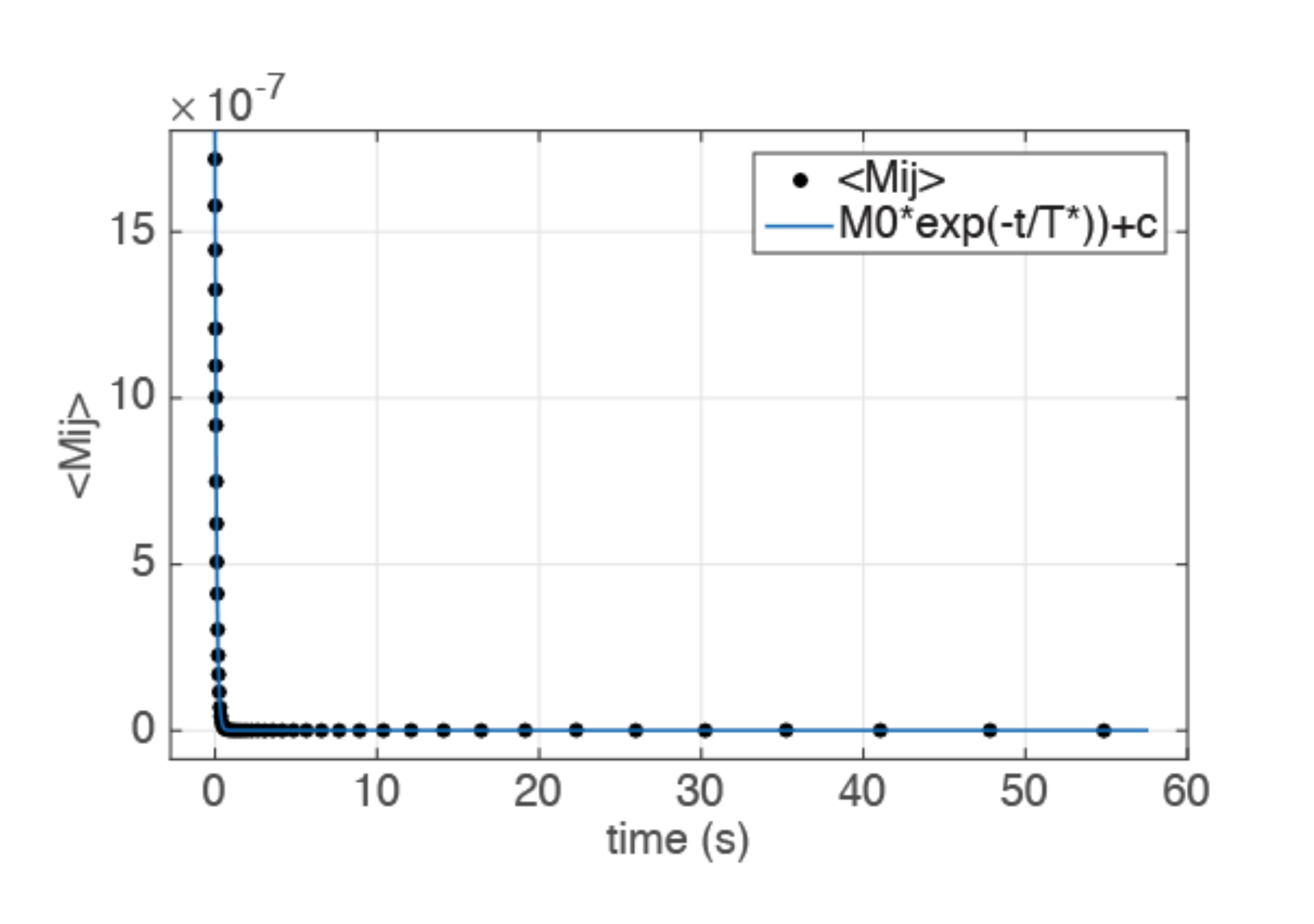}
\caption{The J-coupling relaxation. (\emph{Top}): the quantity $\avg{\tilde{M}_{zz}}$ as a  function of time. (\emph{Bottom}): the quantity $\frac{\avg{\tilde{M}_{xx}}+\avg{\tilde{M}_{xy}}+\avg{\tilde{M}_{xz}}+\avg{\tilde{M}_{yx}}+\avg{\tilde{M}_{yy}}+\avg{\tilde{M}_{yz}}+\avg{\tilde{M}_{zx}}+\avg{\tilde{M}_{zy}}}{8}$ as a function of time.
Notice how small is this coupled magnetisation compared to the uncoupled magnetisations shown in Figs. 3 and 4. Note  also that these ``magnetisations'' are relative to the equilibrium magnetisation, according to Eq. \refe{def_Mj}.}
\label{figs_Mij}
\end{figure}

After running the program in  Eq. \refe{process_tomos}, we plot in Fig. \ref{figs_relax} the trace distance between the states $\rho_0^l$ under the action of the map ${\Phi}(t_k)$ and the rescaled equilibrium state $\frac{1}{M}\rho_{eq}$. We note that the functional form of the curves are  like an exponential decay, as expected.

As a byproduct of the process tomography, the map must give us information about the relaxation times $T_1^{C,H,J}, T_2^{C,H,J}$. We define the quantity
\begin{equation}\label{def_Mj}
\avg{\tilde{M}_{ij}}(t_k) = \frac{\sqrt{ \sum_l \left[ \trace{\tilde{\rho}_k^l\,\sigma_i^H\otimes \sigma_j^C} -  \trace{\tilde{\rho}_{eq}\,\sigma_i^H\otimes \sigma_j^C}  \right]^2 }}{\# l},
\end{equation}
where $\# l$ stands for the number of summed signals. This quantity is the average of the magnetisation in the $i,j$-direction ($i,j \in \{ x, y, z \}$) related to the equilibrium value and it gives us the functional behaviour of the magnetisation during the time. As a first attempt to model  the functional behaviour of Eq. \refe{def_Mj}, one could fit a 
typical exponential  decay  like
\begin{equation}\label{model_fit}
M(t) = M_0 e^{-t/T_*}+c,
\end{equation}
where $T_*$ will be the aimed $T_1^{C,H,J}, T_2^{C,H,J}$. This is of course  a Markovian approximation of the underlying process. 
Notwithstanding, in the top panels of Figs. \ref{figs_MH}-\ref{figs_Mij}, one can see slight deviations of the exponential decay. 
On the one hand, this could be a signature of non-Markovian behaviour, on the other hand, it could also be some artifact caused by the different state preparations
necessary for each observation in time. 
 We should emphasise  that the  tomography  itself is independent of this exponential fit, but this simple  model yields sensible results, in comparison with  previously measured $T_1^{C,H}, T_2^{C,H}$. The newly inferred $T_1^{J}, T_2^{J}$ is also corroborated by  Fig. \ref{unital-trace}, which is independent of the exponential fitting.

{The results presented in Figs. \ref{figs_MH}-\ref{figs_Mij}, and summarised in Table \ref{tab_time} with the estimated errors, were computed using nonlinear least squares fitting with a trust-region algorithm \cite{matlabcurve}. The least squares estimator was chosen to be consistent with the assumption that the experimental errors consist of uniform Gaussian noise,  and also consistent with the $\ell_2$-norm minimization on the observed data in Eq. \refe{process_tomos}.}

The pseudo-pure states corresponding to the canonical states ( $ \ket{\uparrow \uparrow}$, $ \ket{\uparrow \downarrow}$, $\ket{\downarrow \uparrow}$ , $\ket{\downarrow \downarrow}$)
have magnetisation components ($M_{z\mathbb{1}}, M_{\mathbb{1}z}$, $M_{zz}$) only in the z-direction. From these signals, we can retrieve the $T_1^{C,H,J}$, as we can see in the top graphics in Figs. \ref{figs_MH}, \ref{figs_MC}, and \ref{figs_Mij}.

\begin{table}[!ht]
\begin{tabular}{|c|c|c|c|}
\noalign{\smallskip}
\toprule
& $T_1 (s)$ & $T_2 (s)$  \tabularnewline
\midrule
\quad Hydrogen \quad & \quad $6.2 \pm 0.2$ \quad & \quad $0.24 \pm 0.01 $ \quad \tabularnewline
\midrule
\quad Carbon \quad & \quad $7.4 \pm 0.3 $ \quad & \quad $0.192 \pm 0.005 $ \quad \tabularnewline
\midrule
\quad J-coupling \quad & \quad $5.65 \pm 0.07 $ \quad & \quad $0.177 \pm 0.001 $ \quad \tabularnewline
\bottomrule
\noalign{\smallskip}
\end{tabular}
\caption{Relaxation times retrieved from the curve fit in the Figs. \ref{figs_MH}, \ref{figs_MC}, and \ref{figs_Mij} using the model in Eq. \refe{model_fit}.}
\label{tab_time}
\end{table}

The other 16 states have both longitudinal and transversal  magnetisation components ($M_{*\mathbb{1}}, M_{\mathbb{1}*}$, $M_{**}$)  and give us what we need to retrieve the $T_2^{C,H,J}$, as one can see in the bottom graphics in Figs. \ref{figs_MH}, \ref{figs_MC}, and \ref{figs_Mij}.

{From the obtained results, }one can conclude that, if one desires to simulate quantum parameters using the chloroform molecule, the time window where it can be done is $0.177s$, which is the lifetime of the coherent coupling of the spins ($T_2^J$) in the chloroform molecule. 
Otherwise, we can consider the system (partially) relaxed and unable to simulate all parameters properly for a quantum computation.  In Fig.\ref{unital-trace}, we confirm that
$T_2^J$ is the characteristic time window for quantum computation simulations by means of the properties of the quantum process relaxation map. One clearly see in that figure
that the map is almost perfectly unital and trace-preserving up to $\log(T_2^J=0.177s)=-1.731$. After $T_2^J$ the map dramatically departs from been both trace-preserving and
unital, and therefore the normalization problem comes into play.  These results demonstrate the robustness and flexibility of our quantum tomography methodology.

\begin{figure}[!ht]
\centering
\includegraphics[width=15cm]{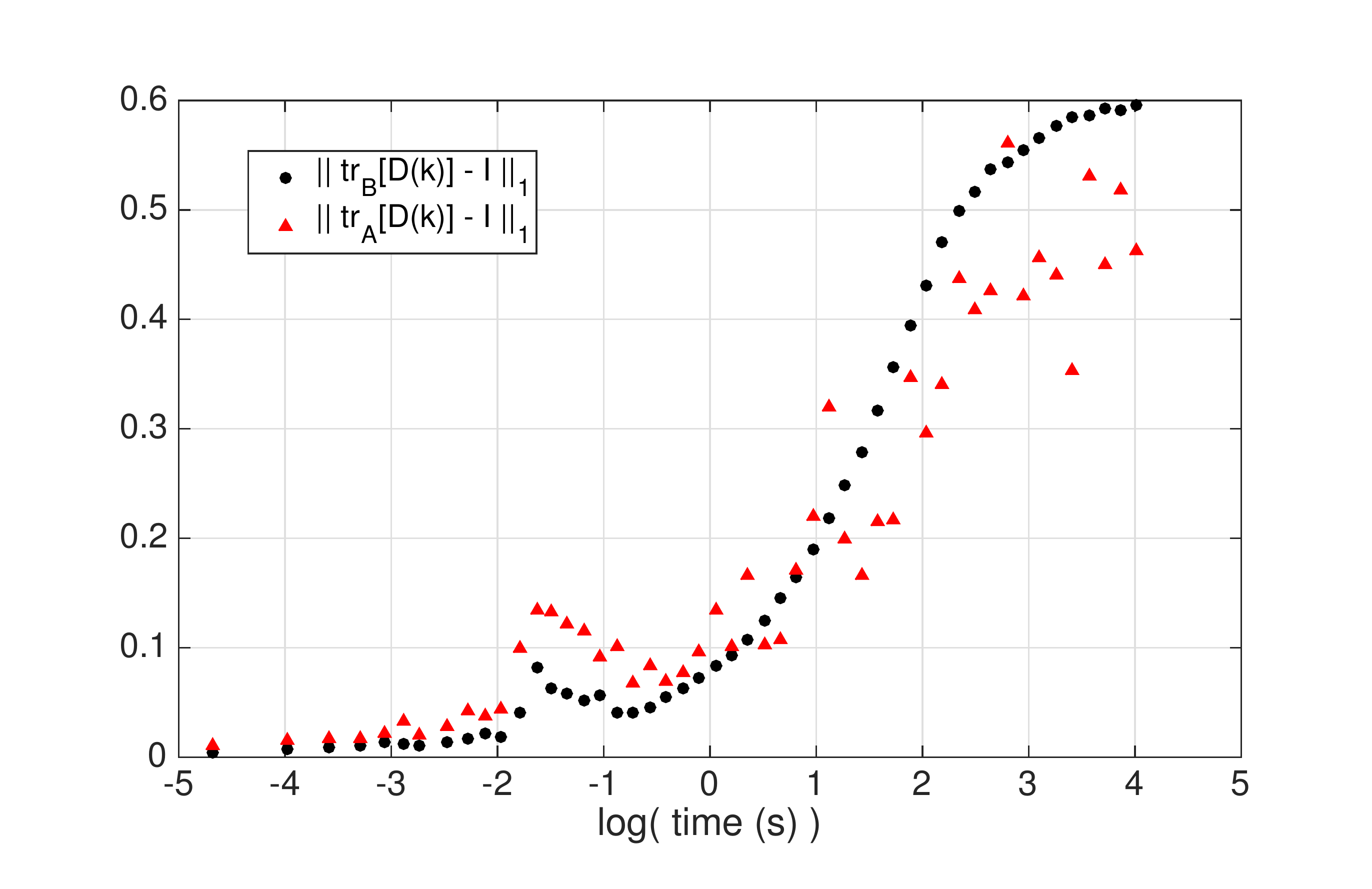}
\caption{Quantification of how much the relaxation process departs from being unital  (\cf Eq. \refe{unital-2}) (black circles),  and trace-preserving
(\cf Eq. \refe{trace-preserving-2}) (red triangles). The dramatic change in the behaviour of the map occurs after $\log{T_2^J} \approx -1.8$. }       
\label{unital-trace}
\end{figure}

\section{Conclusion}
Summarising, we performed a quantum process tomography of the relaxation of  a two-spin system in an NMR liquid-state experiment. As the experimental data correspond to expectation values of traceless operators, the use of informational incomplete quantum tomography  techniques to handle the missing quantum state normalization information was paramount to reconstruct the time dependent non-unital and non trace-preserving 
quantum map. The successful approach we  employed allowed us to introduce two new parameters ($T_1^J, T_2^J$) that are relevant for the characterization of the relaxation
process.  We  learned that the coupled magnetisations have a similar behaviour to the uncoupled ones, but with different characteristic times. 
The quantum process tomography showed  that $T_2^J$ characterizes the time window for the simulation of quantum computation. It can be seen clearly in 
Fig.\ref{unital-trace}. For the simulation of quantum maps for times greater than  $T_2^J$, one has to account for the time dependent normalization of the data, for the 
map associated to the ever present relaxation process cannot be assumed to be unital and trace-preserving. We also introduced an heuristic for the normalization problem,
which consists in assuming the theoretical  equilibrium magnetisation as the reference to compare all the measurements, and the largest trace obtained with the preliminary
state tomography (Eq. \refe{vqtgauss}) as a lower bound for the trace of all states.

\section{Acknowledgements}
We acknowledge Johan Aberg, Paulo H. S. Ribeiro, Lucas C. Céleri and Jefferson G. Filgueiras for  fruitful discussions. 
We are indebted to the anonymous referees for their criticism and suggestions that largely contributed to the final version of this paper.
Financial support by Brazilian agencies CAPES, CNPq, FAPERJ, FAPEMIG, and INCT-IQ (National Institute of Science and Technology for Quantum Information).

\printbibliography

\end{document}